\documentclass[aps,pra,reprint,showpacs,floatfix,amsmath,amssymb,showpacks,nofootinbib]{revtex4-1}

\usepackage{graphicx}
\usepackage{epstopdf}
\usepackage{color}
\usepackage{hyperref}
\usepackage{bm}
\usepackage[left]{lineno}

\usepackage{todonotes}

\newcommand{\Gf}{{\mathcal G}}
\newcommand{\Gfa}{{\mathcal G}_{\text{at}}}
\newcommand{\Gfaa}{\tilde {\mathcal G}_{\text{at}}}
\newcommand{\Ham}{{\mathcal H}}
\newcommand{\Torder}{\text{T}_{\tau} }

\renewcommand{\vec}[1]{\mathbf{#1}}
\newcommand{\ka}{{\vec k}}

\newcommand{\qu}{{\vec q}}

\newcommand{\pe}{{\vec p}}

\newcommand{\ra}{{\vec r}}

\renewcommand{\Re}{\operatorname{Re}}

\newcommand{\Nat}{N_{\text{at}}}

\newcommand{\no}{\noindent}

\begin{document}


\title{One-dimensional Bose-Einstein condensation of photons in a microtube}
\author{Alex J. Kruchkov}
\email{alex.kruchkov@epfl.ch} 
\affiliation{
Laboratory for Quantum Magnetism (LQM), 
\'Ecole Polytechnique F\'ed\'erale de Lausanne (EPFL), Station 3, CH-1015 Lausanne, Switzerland}
     
\date{\today}

\begin{abstract}
This paper introduces a quasiequilibrium one-dimensional Bose-Einstein condensation of photons trapped in a microtube. 
Light modes with a cut-off frequency (a photon's ``mass'') interact through different processes of absorption, emission, and scattering on molecules and  atoms. 
In this paper, we study the conditions for the one-dimensional condensation of light and the role of photon-photon interactions in the system. 
The technique in use is the Matsubara's Green's functions formalism modified for the quasiequilibrium system under study.
\end{abstract}


\maketitle


%
%
%
%

\section*{Introduction}

Light in vacuum is quantized as massless photons, which in equilibrium obey Bose-Einstein statistics. If the photons were massive,  they could break the gauge symmetry under certain conditions and condense to the lowest energy state sharing a single wave function. However, the light nowadays is considered to be massless, so can one expect photons to form the Bose-Einstein condensate? A short answer is yes.

To observe it, the group of Weitz \cite{Klaers2010a,Klaers2010b} used an optical microcavity, where the spectra of light modes have a cut-off due to a geometrical constraint. This cut-off acts as an effective mass for a two-dimensional photon.
Dimensionality here refers to the motional degree of freedom of photons. Although the 2D photons now possess a mass, it is not enough: there is no BEC transition in a uniform two-dimensional system. The condition of uniformity is however broken by the slight curvature of the cavity walls, so the trapped light can be mapped on a 2D field of massive nonrelativistic quasiparticles experiencing harmonic potential \cite{Klaers2010b}, -- the system that is known to undergo the BEC transition. In the experiments \cite{Klaers2010b}, the controllable thermalization process \cite{Klaers2010a,Klaers2011} picks up a single light mode, and small photon losses are compensated by a weak external laser pumping. Therefore, it was shown that the number of photons is conserved in average, and the researchers can keep the system close to its thermodynamical equilibrium. The quasiequilibrium BEC of photons is observed at room temperatures \cite{Klaers2010b,Marelic2015}.

The system being argued to be different from conventional lasers becomes of interest for various theoretical 
\cite{Sobyanin2012,Sobyanin2013,Kruchkov2013,Kirton2013,Leeuw2013, Kruchkov2014,Nyman2014,Leeuw2014a,Leeuw2014b,Leeuw2014c,
Strinati2014,Klaers2012,Weitz2013,Sela2014,Kirton2015} 
and experimental studies 
\cite{Schmitt2014,Marelic2015,Schmitt2015}. 
The growing experimental and theoretical interest in the topic requires broadening the variety of systems for which the condensation of photons could be observed. 
In particular, it was explicitly discussed for dimensionalities $D=2$ (see e.g.
\cite{Klaers2011,Sobyanin2013,Kruchkov2014,Chiao1999})
and in different contexts for $D=3$ (see e.g. \cite{Kruchkov2013,Kuzmin1978,Zeldovich1969}), but never in one dimension.
Therefore, there's a need to complete the study for photons with the one-dimensional degree of freedom.

The theoretical methods applied to the system strongly vary. Some authors are using a phenomenological nonlinear Schr\"{o}dinger equation (Gross-Pitaevskii equation) in different forms \cite{Klaers2010b,Nyman2014,Strinati2014}. The non-interacting $T\ne 0$ theory is also in use in different forms \cite{Klaers2010b,Sobyanin2013,Kruchkov2013,Kruchkov2014}.
The fully off-equilibrium condensate is studied either with an effective kinetic equation with Jaynes-Cummings interaction \cite{Kirton2013,Kirton2015} in the approximation with real-time propagators, or the off-equilibrium $T \ne 0$ Green's function formalism (Schwinger-Keldysh formalism, complex-time propagators, Matsubara's frequencies, etc.)
\cite{Leeuw2013,Leeuw2014a,Leeuw2014b,Leeuw2014c}. 
In my opinion, the Schwinger-Keldysh formalism is the most general and the most powerful approach here.

This paper introduces the one-dimensional quasiequilibrium condensation of photons in a microtube. 
In my opinion, the Matsubara's Green's function formalism is appropriate here for the near-equilibrium system. Of course, the Schwinger-Keldysh formalism may be used here but once the near-equilibrium properties of the system are well understood.
Matsubara's formalism describes the finite-temperature close-to-equilibrium systems, and it is valid for $T \ge T_C$. The main advantage of the approach for this study is that one can calculate the critical parameters of the interacting system.
In this paper, I write down the Hamiltonian, which takes into account one-photon and two-photon processes of interaction with atoms and treat them perturbatively. As a result, I can describe the influence of indirect photon-photon interactions on the critical parameters.

There are some limitations of the model that I use. First, I do not study the thermalization process, as also the time-evolution of the system in general, bounding myself to the steady state only. 
Second, I restrict myself to the (first two) leading corrections to self-energy, which in terms of direct photon-photon interactions, if they were present, would correspond to the Hartree-Fock mean field theory. 
The system under study is a bit more subtle and to obtain these effectively mean-field contributions one need to go to the fourth order in perturbation theory. 
For the same reasons, optical collisions, i.e.,  two-atom mechanical collisions leading to creating of photons,  are not taken into account even though the model in use can do it. 
Other conditions of validity, which shape up the model, are discussed in the main text of the paper as they appear.

This paper is organized into three sections followed by appendices. 
In the first section, the effective mass of light modes is introduced in tubes with varying geometry; 
Then, I discuss the conditions for condensation in 1D and estimate critical parameters.
The second section deals with the interacting theory at $T=T_C$, where the effective Hamiltonian of photon-photon interaction is derived;  this section is the heart of the paper and is organized into subsections for more comfortable reading: it describes the perturbation theory both for the uniform case and for the trapped case. 
The summarizing section is organized more like an outline and discussion, yet the deeper study of the problem is still needed.

%
%
%
%

\section{
\label{Section1}
General idea: light trapping and condensation}

To condense photons in a cavity means to force them going to the lowest-energy thermodynamical state in the system \cite{Klaers2010b,Klaers2011,Sobyanin2013,Kruchkov2014}.
In this section I skip the details of the thermalization process because they are studied sufficiently well \cite{Klaers2010a,Klaers2011,Kirton2013,Leeuw2013}, thus restricting myself to mentioning the three important ideas.
First, the losses of photons are compensated by a weak external pumping, so the number of photons conserves in average.
Second,  the cavity gives a discrete set of light modes with different cut-offs.
Third, it is possible to thermalize one of the modes hence ensuring the single cut-off frequency for all the thermalized photons. 
As a result, supporting only one of the modes, we ascribe the effective mass to a photon as it is described in the first subsection of this section.

 The second challenge for condensing photons in 1D is to choose the shape of the waveguide (microtube) where the condensation is possible. This choice is done in the  context of the non-interacting model in the second subsection. At the end of this section, we discuss the condition for condensation and estimate the critical number of photons for the set of parameters, similar to those from Refs.\cite{Klaers2010a,Klaers2010b,Klaers2011,Marelic2015}.


\subsection{
\label{Subsection1.1}
Light trapping and effective mass of photons}

\begin{figure}[t]
	\includegraphics[width=1.0 \columnwidth]{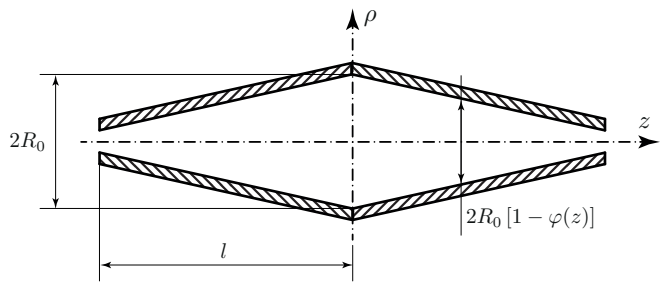}
	\caption{\label{scheme} A scheme of a microtube waveguide for trapping photons. The shape of the tube is determined by the relative deviation $\varphi(z)$ of the inner radius. To ease visual presentation the function $\varphi(z)$ is taken linear.}
\end{figure}

For simplicity, we consider here the waveguides made of a microtube with axial symmetry, as it is sketched on Fig.~\ref{scheme}. The shape of the tube in the general case is given by a rather smooth function $\rho(z)$ (see Fig.~\ref{scheme}). Due to the cylindrical symmetry, a photon's energy $\hbar \omega$ is described by two quantum numbers $k_z$ and $k_\rho$, 

\begin{equation}\begin{split}\begin{gathered}
\label{spectrum}
\hbar \omega (\ka)  = \hbar \tilde c |\ka| = \hbar \tilde c
\left( k_z^2 + k_\rho^2 \right) ^{1/2},
\end{gathered}\end{split}\end{equation}

\no
where $\omega$ is the frequency of a photon with the momentum ${\ka}$ decomposed for convenience into longitudinal $k_z$ and polar $k_\rho$ components. In the microtube the polar wave number $k_\rho$ is  strongly discrete while the longitudinal component $k_z$ can be taken continuous because of the strong inequality $R_0 \ll l$. In the general case, the set of $k_\rho$'s follows from Maxwell equations in the microtube shaped as $\rho (z)$ with the boundary conditions on its walls. For the mirror walls one gets

\begin{equation}\begin{split}\begin{gathered}
\label{k rho}
{k_\rho (z) } = \frac{{{q_{mn}}}}{{\rho \left( z \right)}}, 
\end{gathered}\end{split}\end{equation}

\no
where $q_{mn}$ is the $n$-th root of Bessel funcion of the $m$-th order, $J_{m} \left( q_{mn} \right) = 0$ (see e.g. \cite{Morse Feshbach}).
 The formula \eqref{k rho} was obtained in the approximation of a tube with a slightly changing cross-section radius,

%
%

\begin{equation}\begin{split}\begin{gathered}
\label{form}
\rho \left( z \right) = {R_0}\left[ {1 - \varphi ( z ) } \right],
\ \ \ \ \ 
\varphi \left( z \right) \ll 1,
\end{gathered}\end{split}\end{equation}

\no
where $R_0$ is the radius of the tube at $z=0$, see Fig.~\ref{scheme}.
The dimensionless quantity $\varphi \left( z \right)$ shows the small relative deviation of the tube's radius. 
Due to the the strong demand  $R_{0} \ll l$,  where $l$ is the half-length of the microtube (see the scheme on Fig.~\ref{scheme}), one expects ${\overline{k_z}} \ll k_0$, where ${k_0} = {q_{mn}}/{R_0}$  is the minimal polar wave number. 
As the consequence, the expression \eqref{spectrum} can be asymptotically expanded, 

\begin{equation}\begin{split}\begin{gathered}
\label{expansion}
\hbar \omega     \simeq 
\hbar \tilde c{k_0}
\left(
1 + \frac{{k_z^2}}{{2k_0^2}} + \varphi ( z )
\right).
\end{gathered}\end{split}\end{equation}

\no
This expression can be rewritten in a more intuitive way,

\begin{equation}\begin{split}\begin{gathered}
\label{energy}
\hbar \omega  \simeq  \hbar \omega_{0} + \frac{{{\hbar ^2}k_z^2}}{{2{m^*}}} + 
V(z),
\end{gathered}\end{split}\end{equation}

\no
which reminds a particle with a mass $m^{*}$ and one-dimensional degree of freedom $k_z$ in the field of external potential $V(z)$. In our case, the effective mass of a photon, as follows from comparison between Eqs. \eqref{energy} and \eqref{expansion}, is defined as

\begin{equation}\begin{split}\begin{gathered}
\label{mass}
m^{*} = \frac{\hbar \, q_{mn}}{ \tilde{c} \, R_{0} } ,
\end{gathered}\end{split}\end{equation}

\no
and is related to the cut-off frequency $\omega_{0}$ as $\hbar \omega_{0} =m^{*} \tilde{c}^{2}$, which is a measure of the minimum energy of photons.
The trapping (pseudo)potential, caused by the geometry of the reflective inner surface, is

\begin{equation}\begin{split}\begin{gathered}
\label{kappa}
V(z) =  \frac{q_{mn}\hbar \, \tilde{c} }{R_{0} } \varphi(z).
\end{gathered}\end{split}\end{equation}

\no
Thus, both the effective mass  of a photon inside the tube and the effective  potential take their origin from the specific geometry under consideration or strictly talking, from the $k_\rho \left( z \right)$ component of a photon's wave vector.  

Summarizing the main idea of the subsection, one can say that the system of  photons trapped inside the microtube can be considered as an ensemble of quasiparticles with the mass $m^*$ and the one-dimensional degree of freedom $\kappa = k_z$ placed in the potential $V(z)$. 
 The form of the potential $V(z)$ is determined by the shape of the microtube waveguide $\varphi(z)$.  Therefore, changing the shape of this waveguide, one can change the trapping potential.


\subsection
{\label{Subsection1.2}
Non-interacting theory and critical number of photons}

Noninteracting model is good for a primal estimate. 
In this model, the photons in the microtube are considered as noninteracting particles with the one-dimensional degree of freedom. The total number of photons is given by integrating the Bose-Einstein distribution over the configurational space. 
The condensation condition can be expressed as follows: the chemical potential of photons $\mu_0$ at the critical point reaches the minimum energy of photons in the system, i.e. $\mu_0 = \hbar \omega_0$ (see \cite{Klaers2010b,Kruchkov2014,Kruchkov2013,Sobyanin2013}). 
For the ideal photon gas with the one-dimensional degree of freedom,  the critical number of particles for Bose-Einstein condensation can be estimated in Wigner approximation, 

\begin{equation}\begin{split}\begin{gathered}
\label{Ncrit}
N_{0} 
=
\int 
\frac{ d{k_z} d{z}} {2\pi}
\text{g}^*
\left\{
{\exp\left[ \frac{\hbar ^2 k_z^2/2 m^* +V(z)} { T }  \right] - 1} 
\right\}^{-1},
\end{gathered}\end{split}\end{equation}

\no
where $\text{g}^*$  takes into account the possible degeneracy in photon energy (see e.g. \cite{Kruchkov2013,Kruchkov2014}).
The integral in \eqref{Ncrit} 
may or may not converge, which is a consequence of the Bogoliubov theorem stating, in particular,  that there is no BEC in dimensions below three if the system is uniform. 
However, the presence of the external potential can be considered as nonhomogeneity and the integral in \eqref{Ncrit} is convergent for certain types of potentials; for example, in the case of the one-minimum symmetrical potentials, the singularity is integrable if only the dimensionless  potential $\varphi(z)=V(z)/\hbar \omega_0$ grows slower than a parabolic function,

\begin{equation}\begin{split}\begin{gathered}
\label{condition}
\varphi  \left( z \right)  =    {\left| z / L_0 \right|^\alpha }, \ \ \   \alpha  \in \left( {0,2} \right),   
\end{gathered}\end{split}\end{equation}

\no
where $L_0$ is a parameter in units of length.
Even though one can imagine more sophisticated potentials (for example, multiple-minimum potentials), 
for simplicity, we restrict ourselves to the case of the one-minimum potential of the form \eqref{condition}. 
To calculate the integral \eqref{Ncrit}, we introduce the new variables $\xi$, $\eta$, such as

\begin{equation}\begin{split}\begin{gathered}
\label{new variables}
{\xi ^2} = \frac{{{\hbar ^2}k_z^2}}{{2{m^*}T}},  \ \  \ \ \  {\eta ^\alpha } = \frac{\hbar \omega_0 }{T}{\left| {\frac{z}{{{L_0}}}} \right|^\alpha },  
\end{gathered}\end{split}\end{equation}

\no
and after some hackneyed algebra obtain the expression for the critical number of photons in the system to observe the phase transition at the temperature $T$,

\begin{equation}\begin{split}\begin{gathered}
\label{N_C}
N_0 (T; \alpha)
=
\text{g}^* \frac{ 2 \sqrt{2}  }{\pi   } 
\frac{ L_0 \omega_0}{\tilde c}
\left( \frac{T}{\hbar \omega_0}  \right)^{\frac{1}{2}+\frac{1}{\alpha}}
I\left( \alpha  \right), 
\end{gathered}\end{split}\end{equation}

\no
where we have introduced the dimensionless normalization integral, 

\begin{equation}\begin{split}\begin{gathered}
\label{normalization integral}
I\left( \alpha  \right) = \int\limits_0^\infty  {\int\limits_0^\infty  {\frac{{d\xi d\eta }}{{\exp \left( {{\xi ^2} + {\eta ^\alpha }} \right) - 1}}} }. 
\end{gathered}\end{split}\end{equation}

%
\begin{figure}
	\includegraphics[width=0.9\columnwidth]{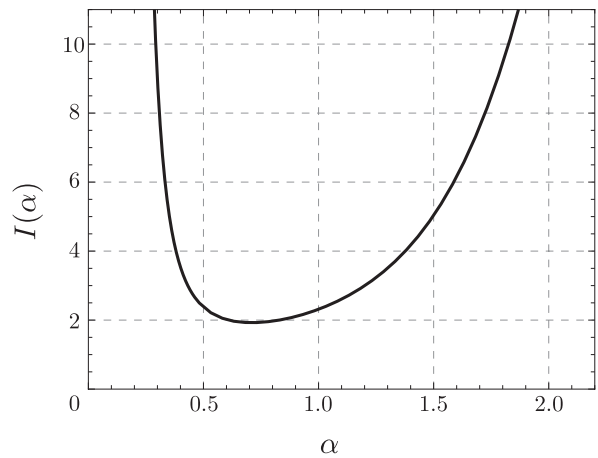}
	\caption{\label{normintegral}
Normalization integral
  $I(\alpha)$ as a function of 
the trapping parameter $\alpha$, $V(z) \propto |z|^\alpha$. The local minimum is situated at $\alpha_{\text{min}} = 0.71$.  The two asymptotes	(not shown) are $I(\alpha) \to + \infty$ as $\alpha \to 0$ and $\alpha \to 2$, limiting the region 
of desirable trapping parameters to
$\alpha \in (0,2)$, where the condensation of an ideal gas of photons is allowed in one dimension.
	}
	\label{tube}
\end{figure}
%

\no
The normalization integral remains finite while the trapping parameter is  $\alpha \in \left( 0, 2 \right)$. The dependence $ I \left( \alpha \right)$ is shown
  in Fig.~\ref{normintegral}. Nameworthy, there is a minimum at the trapping parameter $\alpha_{\text{min}} = 0.71$, $ I \left( \alpha_{\text{min}} \right) = 1.9$. 
However, a more fruitful trapping parameter is
 $\alpha =1$, i.e. $\varphi (z) \propto \left| z \right| $, where the value of the normalization integral (as also other quantities of non-interacting and interacting theory) can be calculated analytically,  $I(1) = \Gamma(3/2) \zeta(3/2)$. In this case, the expression \eqref{N_C} for the critical number of photons simplifies,

\begin{equation}\begin{split}\begin{gathered}
\label{N crit linear 1}
N_0 = 
\sqrt{\frac{2}{\pi}} 
\, \text{g}^{*} \zeta \left(3/2 \right) 
\frac{L_0 \omega_0}{\tilde c}
\left(\frac{T}{\hbar \omega_0}  \right)^{3/2}.
\end{gathered}\end{split}\end{equation}

\no
Taking into account the explicit expression for $\omega_0$, and $\zeta (3/2) \approx 2.6 $, 
$\text g^{*} \approx 3$
the formula \eqref{N crit linear 1} can be simplified and given in terms of direct experimental parameters, 

\begin{equation}\begin{split}\begin{gathered}
\label{N crit linear 2}
N_0 \approx 
\left( \frac{T}{\hbar \tilde c k_{\Lambda}} \right)^{3/2},
\ \ \ \ \
k_{\Lambda} = 
\left(
\frac{q_{mn}}{4 \pi^2 R_0 L_0^2 } 
\right)^{1/3} .
\end{gathered}\end{split}\end{equation}

\no
The formula \eqref{N crit linear 2} defines the critical number of photons in the tube with V-like trapping profile, $\varphi (z ) \propto \left| z \right|$. Such a biconical waveguide can be indeed manufactured \cite{Vogl2011}.
It is remarkable that the $T^{3/2}$ dependence, as in the formula \eqref{N crit linear 2} for 1D gas of particles under  V-like  potential, also holds for 3D uniform gas of bosons.  This similarity rises from the composition of the Wigner integral \eqref{Ncrit}. 

An estimate for the biconical waveguide gives following. For definiteness, we take the lowest Bessel root $q_{01} \approx 2.4$. The radius of the waveguide should be such to ensure closeness of cut-off and the atomic transition frequencies, giving $R_0 \approx q_{01} \lambdabar_{\text{at}}$, where $\lambdabar_{\text{at}}$ is atomic transition wavelength reduced by $2 \pi$.
Taking now for estimate 
$\lambdabar_{\text{at}} \sim 10^{-6} \, \text{m} $, $\tilde c \approx 2.2 \cdot 10^8 \, \text{m/s}$, $L_0 \sim 10^{-2} \, \text{m}$, and the room temperature $T=300 \,  \text{K}$, the threshold number of photons to trig condensation is $N_0 \sim 10^4$, which is even smaller than the one reported for the 2D condensation, $N_0 \sim 10^5$. Thus, one may conclude that even at room temperatures the one-dimensional condensation of photons is possible.

%
%
%
%

\section{\label{Section2}
Interacting theory}

As we have seen in the previous section, the non-interacting model is good. It works for a range of potentials and predicts the BEC transition of photons in one dimension. The non-interacting model would be exact for photons in a vacuum where their scattering cross-section is negligible. However, the photons in the system under study do interact with each other, be it indirectly.  These interactions arise from the multiple acts of scattering, absorption and emission of photons. One can classify all these processes by the number of photons involved in a single act, and then construct a hierarchy of irreducible acts (the events, which cannot be represented as a product of two different acts). This hierarchy defines the form of an effective interacting Hamiltonian, which then is treated perturbatively.

To build up the perturbation theory, first I write down the Hamiltonian of the irreducible interactions, which is done in the first subsection. In the second subsection, the renormalized Green's functions of photons are derived for the uniform (non-trapped) case. The effect of the trapping potential is considered in the third subsection. Finally, the last subsection gives contributions for all the one-photon and two-photon processes.
This section is rich on physics and intended to be longer.


\subsection{
\label{Subsection2.1}
Interacting Hamiltonian and observables}

The interacting Hamiltonian $\Ham$ should include processes of absorption, emission and scattering of photons on atoms. It can be naturally written in a secondly-quantized form. For this, we introduce the operators of creation
$\phi^{\dag}_{\kappa_\ka}$ and annihilation $\phi^{}_{\kappa_\ka}$ of a photon as a massive quasiparticle with the one-dimensional degree of freedom 
$\kappa_\ka =  ({\ka^2-k_{0}^2})^{1/2} \equiv \kappa$. 
In the cylindrical microtube, as we have already shown, these quasiparticles have the quadratic dispersion law $\hbar \omega_\kappa = \hbar \omega_0 + \hbar^2 \kappa^2/2m^*$ and are placed in the field of  trapping potential. Consequently, the secondly quantized Hamiltonian is given in a general form as

\begin{equation}\begin{split}\begin{gathered}
\Ham = \sum_{\ka} \hbar \omega_{\kappa_\ka}  
\phi^{\dag}_{\kappa_\ka} \phi^{}_{\kappa_\ka}
+\sum_{\ka,\qu} V_{\qu}  \, 
\phi^{\dag}_{\kappa_{\ka+\qu}} \phi^{}_{\kappa_\ka}+\Ham_{I},
\end{gathered}\end{split}\end{equation} 

\no
where $V_{\qu} $ is the Fourier-transform of the trapping potential $V(z)$ and $\Ham_{I}$ reflects the photon-atom interactions.
Here are some examples of the elementary interaction acts: A photon can be absorbed by an atom in the ground state; A photon can be emitted by an atom in an excited state; A photon can be scattered by an atom (or in secondly-quantized language destroyed and then created again). 
To describe these events quantomechanically, 
an adequate atomic model in use is the so-called two-level model, where an atom can be in two states: the ground state $\left| {E_{\sigma_1}(\pe)} \right\rangle$ and the excited state $\left|{E_{\sigma_2}(\pe) }\right\rangle$. The validity of this model for this study is satisfied by the two reasons mainly: First, the cut-off of photons is set closer to the chosen atomic transition, $\hbar \omega_0 \approx \hbar \omega_{\text{at}}$, and the other transitions are energetically separated; Second, the number of photons is small in comparison  to the number of atoms, $N_\phi \ll N_{\text{at}}$. As a consequence, the probability of exciting the higher states is strongly suppressed and can be neglected in the main approximation.

%
\begin{figure*}
	\includegraphics[width=0.98 \textwidth]{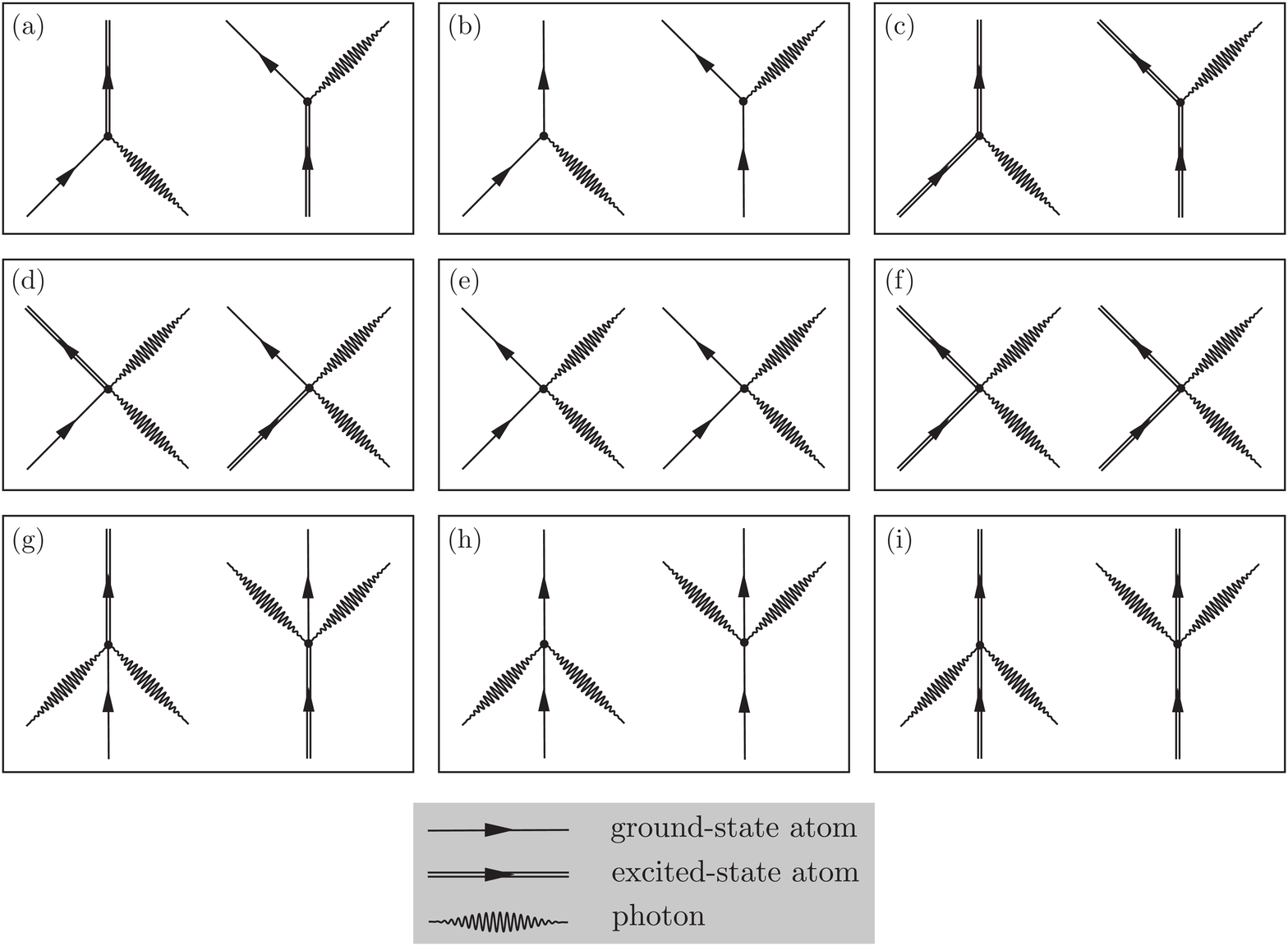}
	\caption{\label{diagrams} 
	Hierarchy of the interaction processes: a)-c) 0 One-photon processes, d-i) Two-photon processes. Notations: single line ground-state atom; double line -excited-state atom; curly line - photon. Time flows from up to top of a diagram.}
\end{figure*}
%

We introduce the operators of creation and annihilation of atoms in the ground state $a^{\dag}_{\pe}$, $a^{}_{\pe}$ and in the excited state $\tilde a^{\dag}_{\pe}$, $\tilde a^{}_{\pe}$, where $\pe$'s label the atomic momenta.
The photon-atom interactions are described now as all the possible combinations of $\phi$'s, $a$'s and $\tilde a$'s (times the complex-valued coupling vertices), and the number of these combinations is, in principle, infinite. The good news here is that one can, for example, build up a hierarchy of irreducible processes based on the number of photons involved in a process. ``Irreducible'' stands here for a process which cannot be decomposed into two (or more) simpler processes. 

The important one-photon and two-photon processes are sketched in Fig.~\ref{diagrams}.
For example, the left diagram in the subfigure (a) of the Fig.~\ref{diagrams} shows a simplest one-photon process: a ground-state atom absorbs a photon and goes to the excited state. In the conjugated process, shown in the right diagram of the subfigure (a),  the excited atom emits a photon and goes to the ground state. 
Hence, the interacting Hamiltonian can be expressed as

\begin{equation}\begin{split}\begin{gathered}
\label{hierarchy}
\Ham_{I} 
=
\Ham_{I}^{11}
+ 
\Ham_{I}^{12}
+
\Ham_{I}^{13}
+ 
\Ham_{I}^{21}
+ 
\Ham_{I}^{22}
+ 
\Ham_{I}^{23}
+\dots ,
\end{gathered}\end{split}\end{equation}

\no
where the one-photon processes are described by

\begin{equation}\begin{split}\begin{gathered}
\label{one-photon}
\Ham_{I}^{11} = 
\frac{1}{\sqrt{\Nat}}
\sum_{\pe, \ka}
   \Gamma^{11}_{\ka}
  \ \tilde a^{\dag} _{\pe + \ka} \, a^{}_{\pe} \, \phi^{}_{\kappa_\ka} +H.c. ,
\\
{\cal H}_{I}^{12}= 
\frac{1}{\sqrt{\Nat}}
\sum_{\pe, \ka}
  \Gamma^{12}_{\ka} 
 \       a_{\pe + \ka}^{\dag} \, a^{}_{\pe} \, \phi^{}_{\kappa_\ka}  + H.c. ,
\\
{\cal H}_{I}^{13}= 
\frac{1}{\sqrt{\Nat}}
\sum_{\pe, \ka}
  \Gamma^{13}_{\ka}
   \ \tilde a_{\pe + \ka}^{\dag} \, \tilde a^{}_{\pe} \, \phi^{}_{\kappa_\ka}   + H.c. ,
\end{gathered}\end{split}\end{equation}

\no
the two-photon processes are described by

\begin{equation}\begin{split}\begin{gathered}
\label{two-photon}
\Ham_{I}^{21} = 
\frac{1}{\sqrt{\Nat}}
\sum_{\pe, \ka,\qu}
 \Gamma^{21}_{\ka \qu}  \
 \tilde a^{\dag} _{\pe + \ka - \qu}  \, \phi^{\dag}_{\kappa_\qu} \,
  a^{}_{\pe} \, 
  \phi^{}_{\kappa_\ka}   + H.c.,
\\
\Ham_{I}^{22} = 
\frac{1}{\sqrt{\Nat}}
\sum_{\pe, \ka,\qu}
\Gamma^{22}_{\ka \qu}  \
a^{\dag} _{\pe + \ka-\qu}  \, \phi^{\dag}_{\kappa_\qu} \,
 a^{}_{\pe} \, \phi^{}_{\kappa_\ka}   + H.c., 
\\
\Ham_{I}^{23} =
\frac{1}{\sqrt{\Nat}}
 \sum_{\pe, \ka,\qu}
\Gamma^{23}_{\ka \qu}  \
\tilde a^{\dag} _{\pe + \ka-\qu}  \, \phi^{\dag}_{\kappa_\qu} \,
 \tilde a^{}_{\pe} \, \phi^{}_{\kappa_\ka}     + H.c., 
 \\
\Ham_{I}^{24} = 
\frac{1}{\sqrt{\Nat}}
\sum_{\pe, \ka,\qu}
 \Gamma^{24}_{\ka \qu}  \
 \tilde a^{\dag} _{\pe + \ka + \qu}  \, \phi^{}_{\kappa_\qu} \,
  a^{}_{\pe} \, \phi^{}_{\kappa_\ka}   + H.c.,
  \\
\Ham_{I}^{25} = 
\frac{1}{\sqrt{\Nat}}
\sum_{\pe, \ka,\qu}
\Gamma^{25}_{\ka \qu}  \
a^{\dag} _{\pe + \ka+\qu}  \, \phi^{}_{\kappa_\qu} \,
 a^{}_{\pe} \, \phi^{}_{\kappa_\ka}   + H.c., 
 \\
\Ham_{I}^{26} =
\frac{1}{\sqrt{\Nat}}
 \sum_{\pe, \ka,\qu}
\Gamma^{26}_{\ka \qu}  \
\tilde a^{\dag} _{\pe + \ka + \qu}  \, \phi^{}_{\kappa_\qu} \,
 \tilde a^{}_{\pe} \, \phi^{}_{\kappa_\ka}     + H.c., 
\end{gathered}\end{split}\end{equation}

\no
and so on. For beauty, I adopt $\hbar=1$ in indices labelling, for example 
$\tilde a^{\dag} _{\pe + \ka} \equiv \tilde a^{\dag} _{\pe + \hbar \ka}$ 
reads as an excitation of an atom with the momentum $\pe +\hbar \ka $. In general, I tend to keep $\pe$ and $\pe'$ for atomic momenta and $\ka$ and $\qu$ for photon wave vectors, so it is easy to distinguish. The coupling parameters $\Gamma_{\ka}$ should be also read as $\Gamma_{\ka} = \Gamma( \omega_{\ka})$ due to their scalar nature. 
In the present paper, we neglect the contributions from optical collisions, i.e. the processes of form 
$a^{\dag}_{\pe_1} \,  a^{\dag}_{\pe_1} \, \phi^{\dag}_{\kappa_\qu} \, a^{}_{\pe'_1} \, a^{}_{\pe'_2}$
  (with $\qu = \pe_1+\pe_2 - \pe'_1 - \pe'_2$) and others, even though these processes can be taken into account in this model by writing down their one-photon and two-photon Hamiltonians in the secondly-quantized form. 
For simplicity here we consider $\hbar = 1$ until the end of the section, where it is restored.

We define the Matsubara's Green's function of a photon as

\begin{equation}\begin{split}\begin{gathered}
\Gf ( \kappa, \tau, \tau_0) = - \left \langle \, 
\Torder \, 
\phi_{\kappa} (\tau)  \, \phi^{\dag}_{\kappa} (\tau_0)
\right \rangle_{\text{th}},
\end{gathered}\end{split}\end{equation}

\no
where all the operators are in the (imaginary-time) Heisenberg representation, and $\Torder$ stands for Matsubara time ordering.
We introduce now Fourier-transformed Green's functions,

\begin{equation}\begin{split}\begin{gathered}
\label{Fourier transf}
 \Gf(\kappa, i \omega_n) = 
\int \limits_{0}^{\beta} d\tau e^{i \omega_n \tau} 
\Gf (\kappa, \tau) ,
\end{gathered}\end{split}\end{equation}

\no
where $i \omega_n$ are Matsubara frequencies, which are of discrete nature, $\omega_n = 2 \pi n/\beta$ for bosons, and $\beta=1/T$ is the inversed temperature as usually.

We also introduce the atomic Green's functions as

\begin{equation}\begin{split}\begin{gathered}
 \Gfa ( \pe, \tau, \tau_0) = - \left \langle \,
\Torder \, 
a_{\pe} (\tau)  \, a^{\dag}_{\pe} (\tau_0)
\right \rangle _{\text{th}},
\\
\Gfaa ( \pe, \tau, \tau_0) = - \left \langle  \, 
\Torder \, 
\tilde a_{\pe} (\tau)  \, \tilde a^{\dag}_{\pe} (\tau_0)
\right \rangle_{\text{th}}.
\end{gathered}\end{split}\end{equation}

\no
Some properties of the atomic Green's functions are given in Appendix A.

Matsubara Green's functions link observables at finite temperatures with the equal-time response of quantum operators. For instance, the occupation number of photons with one-dimensional degree of freedom $\kappa$ is given by

\begin{equation}\begin{split}\begin{gathered}
\label{occ number photons}
f_{\kappa} = 
\langle \phi^{\dag}_{\kappa} \phi_{\kappa} \rangle
=
- \lim_{\tau \to  0^{-} } \Gf (\kappa,\tau, 0),
\end{gathered}\end{split}\end{equation}

\no
and similar expression are for the occupation number of atoms in the three-dimensional reciprocal space $\pe$,  

\begin{equation}\begin{split}\begin{gathered}
n_{\pe} = 
\langle a^{\dag}_{\pe} a^{}_{\pe} \rangle
=
- \lim_{\tau \to  0^{-} } \Gfa (\pe,\tau, 0),
\\
\tilde n_{\pe} =
\langle \tilde a^{\dag}_{\pe} \tilde a^{}_{\pe} \rangle
=
 - \lim_{\tau \to  0^{-} } \Gfaa (\pe,\tau, 0).
\end{gathered}\end{split}\end{equation}

\no
One may notice that the photon occupation number $f_{\kappa}$ and atomic occupation numbers $n_{\pe}$, $\tilde n_{\pe}$ are introduced by different literals. This is intended. First, I wish to distinguish between them two without adding additional indices. Second, the two quantities are of different units, so they represent different physical notions.


\subsection{Perturbation theory in a uniform medium}

In this subsection, we derive the renormalized photon propagator and the corresponding self-energies in the absence of the trapping potential. 
In the Matsubara's formalism, the perturbed Green's function is given by a series expansion 

\begin{equation}\begin{split}\begin{gathered}
\label{perturbation series}
\Gf(\kappa, \tau, \tau_0 ) = - 
\sum \limits_{n=0}^{\infty} 
\frac{(-1)^n}{n!}
\int \limits_{0}^{\beta} d\tau_1 ... \int \limits_{0}^{\beta}
d\tau_n \  
\\
\times
{ \left \langle  
\Torder   \ 
 \phi^{}_{\kappa}(\tau)
\  \phi^{\dag}_{\kappa}(\tau_0)  
 \ \Ham_{I} (\tau_1)  ...  \Ham_{I} (\tau_n)  \, 
\right \rangle}_{0} ,
\end{gathered}\end{split}\end{equation}

\no
where the thermal averaging includes only connected diagrams, and $0$ stands for the non-interacting eigenstates. 
To simplify the discussion, in this subsection I 
consider a single process only, namely the first one,  

\begin{equation}\begin{split}\begin{gathered}
\label{11 Ham}
\Ham_{I}^{11} = 
\frac{1}{\sqrt{\Nat}}
\sum_{\pe, \ka}
\Gamma^{11}_{\ka} \, \tilde a^{\dag} _{\pe + \ka}  \, 
a^{}_{\pe} \, \phi^{}_{\kappa_\ka}   
+ \Gamma^{11*}_{\ka}  \,   \phi^{\dag}_{\kappa_\ka} \,  a^{\dag}_{\pe} \, \tilde a^{} _{\pe + \ka} .
\end{gathered}\end{split}\end{equation} 

\no
This choice is not arbitrary. Indeed, it is $\Ham_{I}^{11}$ that gives one of the most significant contributions to the self-energy. At the end of this section, the contributions from all the other one-photon and two-photon processes are taken into account.  
To simplify the relevant formulas further, I drop in this subsection all the ''$11$'' superscripts of expression \eqref{11 Ham}.


According to the formula \eqref{perturbation series},
the first non-vanishing correction that is given due to acts of absorption and re-emission is given by the first mean-field correction, is

\begin{equation}\begin{split}\begin{gathered}
\delta \Gf^{(1)}(\kappa, \tau) = - 
\frac{1}{2!}
\int \limits_{0}^{\beta} d\tau_1 \int \limits_{0}^{\beta}
d\tau_2  
\\
\times
{ \left \langle  
\Torder   \, 
\phi^{}_{\kappa}(\tau)
\,  \phi^{\dag}_{\kappa}(0)  
  \, \Ham_{I} (\tau_1)  \,
 \Ham_{I} (\tau_2)  \, 
\right \rangle}_0 .
\end{gathered}\end{split}\end{equation}

\no
Using the explicit expression \eqref{11 Ham}, and decoupling the time-ordered average by using the Wick's theorem, the contributions of the connected diagrams are determined as

\begin{equation}\begin{split}\begin{gathered}
\label{G1}
\delta \Gf^{(1)}(\kappa_\ka, \tau) = 
\frac{|\Gamma_{\ka}|^2}{\Nat}
\int \limits_{0}^{\beta} d\tau_1 \int \limits_{0}^{\beta}
d\tau_2 
\
\Gf (\kappa_{\ka}, \tau - \tau_2 ) 
\,
\Gf (\kappa_{\ka}, \tau_1 ) 
\\
\times
\sum_{\pe} 
\Gfa (\pe,\tau_1 - \tau_2)
\,
 \Gfaa (\pe + \ka, \tau_2 - \tau_1) .
\end{gathered}\end{split}\end{equation}

\no
Going to the $i \omega_n$-representation of formula \eqref{G1}, 
 one  shows  the non-vanishing contribution corresponds to $\tau_1 = \tau_2$. 
For the non-degenerate ensemble of atoms, the product of the atomic propagators is given as
 $\Gf_{at} (\pe,\tau_1 - \tau_2) \tilde \Gf_{at} (\pe + \ka, \tau_2 - \tau_1) = n_{\pe} \tilde n_{\pe + \ka}$. 
Thus, after Fourier-transforming Eq.\eqref{G1}, one  obtains the first renormalization to the Green's function

\begin{equation}\begin{split}\begin{gathered}
\delta \Gf^{(1)}(\kappa,i \omega_n)   = 
\Sigma^{(1)} (\kappa, i \omega_n)  \,   \Gf^2_{0}(\kappa,i \omega_{n}),
\end{gathered}\end{split}\end{equation} 

\no
where the first (on-shell) self-energy $\Sigma^{(1)} (\kappa, i \omega_n) = \Sigma^{(1)}_\kappa$ is contributed by one-photon emission/absorption,

\begin{equation}\begin{split}\begin{gathered}
\Sigma^{(1)}_{\kappa_\ka} = 
\frac{|\Gamma_{\ka}|^2}{\Nat}
\sum_{\pe} n_{\pe} \tilde n_{\pe + \ka}.
\end{gathered}\end{split}\end{equation}

\no
To calculate the density of particles, we go back to the $\tau$-representation, namely considering the perturbed Green's function  as

\begin{equation}\begin{split}\begin{gathered}
\Gf(\kappa,\tau )   
\approx  \frac{1}{\beta}   \sum_{i \omega_n} e^{- i \omega_n \tau }   \Gf_{0}(\kappa,i \omega_{n}  )
\\
+  \frac{1}{\beta}   \sum_{i \omega_n} e^{- i \omega_n \tau }   \Gf^2_{0}(\kappa,i \omega_{n}  )  \Sigma^{(1)}_{\kappa} .
\end{gathered}\end{split}\end{equation}

\no
As a result, the occupation number of photons, renormalized due to interactions, is obtained through the  $\tau \to 0^{-}$ limit, in the main order one obtains

\begin{equation}\begin{split}\begin{gathered}
\label{first order occ number}
f_{\kappa}
=   
- \frac{1}{\beta}   \sum_{i \omega_n}   \Gf_{0}(\kappa,i \omega_{n}  ) 
 -   \frac{1}{\beta}   \sum_{i \omega_n}   
  \Gf^{2}_{0}(\kappa,i \omega_{n}) \,  \Sigma^{(1)}_{\kappa}
 \\
= 
-  \frac{1}{\beta}   \sum_{i \omega_n}  
\frac{1}{i \omega_n - \omega_{\kappa}}
 - 
 \frac{1}{\beta}   \sum_{i \omega_n}  
\frac{\Sigma^{(1)}_{\ka}   }{(i \omega_n - \omega_{\kappa})^2}.
\end{gathered}\end{split}\end{equation}

\no
Using now the Matsubara frequencies summation rules
(see Ref.\cite{Mahan}), one derives

\begin{equation}\begin{split}\begin{gathered}
\label{first order occ number 2}
f_{\kappa} = 
\frac{1}{e^{\beta \omega_{\kappa}}-1}
- 
 \frac{ \beta \Sigma_{\kappa}^{(1)}  e^{\beta \omega_{\kappa}} }{(e^{\beta \omega_{\kappa} }-1)^2 }.
\end{gathered}\end{split}\end{equation}

\no
So far the existence of the non-zero chemical potential $\mu$  was for simplicity avoided in \eqref{first order occ number 2}, but it is easily included switching $\omega_{\kappa} \to \omega_{\kappa} - \mu$ in the Green's functions of photons. For the non-zero chemical potential, the renormalized occupation number in the first order is

\begin{equation}\begin{split}\begin{gathered}
\label{first occ number}
f (\omega_{\kappa} +\Sigma_{\kappa}^{(1)} - \mu) \approx 
\frac{1}{e^{\beta (\omega_{\kappa}- \mu)}-1}
- 
 \frac{ \beta \Sigma_{\kappa}^{(1)}  e^{\beta (\omega_{\kappa}- \mu)} }{\left(e^{\beta (\omega_{\kappa}- \mu) }-1 \right)^2 }.
\end{gathered}\end{split}\end{equation}

\no
 The first term in \eqref{first occ number} is intuitively understandable as it describes the occupation number for the non-interacting gas of photons while the second term gives the first interacting correction. We still keep here $\hbar=1$.


The next non-vanishing correction is given by the fourth order of the perturbation theory,

\begin{equation}\begin{split}\begin{gathered}
\label{Gf2}
\delta \Gf^{(2)}(\kappa, \tau) = -  \frac{1}{4!}
\int \limits_{0}^{\beta}  \int \limits_{0}^{\beta} \int \limits_{0}^{\beta}  \int \limits_{0}^{\beta}
d\tau_1 d\tau_2 d\tau_3 d\tau_4  
\\
\times
{ \left \langle  
\Torder   \ 
\phi^{}_{\kappa}(\tau)
  \phi^{\dag}_{\kappa}(0)  
   \Ham_{I} (\tau_1)  
 \Ham_{I} (\tau_2) 
 \Ham_{I} (\tau_3)   
 \Ham_{I} (\tau_4)  
\right \rangle}.
\end{gathered}\end{split}\end{equation}

\no
The unique connected diagrams are given, for example, by $\tau_4 = \tau_1$, $\tau_3 = \tau_2$. Decoupling operators with the use of Wick's theorem, one finds the quantity under averaging  $\langle \Torder \dots \rangle$ in  Eq.\eqref{Gf2} containing the following combination of the atomic Green's functions,

\begin{equation}\begin{split}\begin{gathered}
\Gfa (\pe, \tau_{12}) 
\Gfa(\pe', \tau_{21})  
\Gfaa (\pe + \ka, \tau_{21}) 
\Gfaa (\pe' + \qu, \tau_{12})
\\
+
\Gfa (\pe, 0 )  
\Gfa(\pe', 0)  
\Gfaa (\pe + \ka, 0) 
\Gfaa (\pe' + \ka, 0)
\\
+
\Gfa (\pe, \tau_{12}) 
\Gfa(\pe+\ka - \qu, \tau_{21})  
\Gfaa^2(\pe + \ka, 0)
\\
+
\Gfa^2 (\pe, 0 )
\Gfaa (\pe + \ka, \tau_{21}) 
\Gfaa (\pe + \qu, \tau_{12}),
\end{gathered}\end{split}\end{equation}

\no
where we have introduced $\tau_{12} = \tau_1 - \tau_2$ and $\tau_{21} = \tau_2 - \tau_1$ for brevity.
Using the properties of the Matsubara's Green's functions, this expression can be transformed to

\begin{equation}\begin{split}\begin{gathered}
\Gfa (\pe, \tau_{12})  
\Gfa(\pe', \tau_{21})   
\Gfaa (\pe + \ka, \tau_{21})
\Gfaa (\pe' + \qu, \tau_{12})
\\
\times
\left(
1 +  \delta_{\pe, \pe'} + \delta_{\ka, \qu}+ \delta_{\pe + \ka, \pe' + \qu}
\right).
\end{gathered}\end{split}\end{equation}

\no
This, in turn, gives the second perturbative correction to the photon Green's function as

\begin{equation}\begin{split}\begin{gathered}
\delta \Gf^{(2)}(\kappa_\ka, \tau) =  
\frac{|\Gamma_{\ka}|^2}{\Nat}
\sum_{\qu} 
\frac{ |\Gamma_{\qu}|^2} {\Nat} 
 \sum_{\pe,\pe'}
\int \limits_{0}^{\beta}  \int \limits_{0}^{\beta} 
d\tau_1 d\tau_2  \, 
\\
\times 
\Gf \left(\kappa_\ka, \tau- \tau_2 \right)  \, 
\Gf \left( \kappa_\qu, \tau_2 - \tau_1 \right) \,
\Gf \left(\kappa_\ka, \tau_1 \right)
\\
\times \Gfa (\pe, \tau_1 - \tau_2)  
\Gfa(\pe', \tau_2 - \tau_1) 
\Gfaa (\pe + \ka, \tau_2 - \tau_1) 
\\
\times \Gfaa (\pe' + \qu, \tau_1 - \tau_2) 
\left(
1 +  \delta_{\pe, \pe'} + \delta_{\ka, \qu}+ \delta_{\pe + \ka, \pe' + \qu}
\right).
\end{gathered}\end{split}\end{equation}

\no
Now, we use again the condition that the atomic ensemble is  non-degenerate, yielding a quasiclassical propagator
 $
\Gfa (\pe, \tau) =  - e^{- E_{\pe} \tau} n_{\pe}
$,
and a similar expression for $ \Gfaa (\pe, \tau)$. 
Therefore, one obtains

\begin{equation}\begin{split} \begin{gathered}
\label{prod at}
\Gfa (\pe, \tau_{12}) 
\Gfaa (\pe + \ka, \tau_{21}) 
\Gfa(\pe', \tau_{21})  
\Gfaa (\pe' + \qu, \tau_{12})
\\
=
n_{\pe} \tilde n_{\pe+ \ka}  \ n_{\pe'} \tilde n_{\pe'+ \qu} \ 
e^{(\omega_{\qu} -  \omega_{\ka}) \tau_{21} },
\end{gathered} \end{split}\end{equation}

\no
where we have used the energy conservation laws, $ E_{\pe} + \omega_{\ka} = E_{\pe+\ka} + \Delta$, with $\Delta = \omega_{\text{at}}$ is the energy distance between the ground state  and the excited state, and a similar expression for $\qu$. 
The vector labelling of photon energies is used here to emphasize that this expression \eqref{prod at} holds in general. 
For definiteness, we consider $\tau_2 > \tau_1$ in the present calculation.
Next step is proceeding to the Matsubara frequencies, 
and we are doing the Fourier transform to Matsubara frequencies,

\begin{equation}\begin{split} \begin{gathered}
\delta \Gf ^{(2)}(\kappa_\ka, i \omega_n) =  
\frac{|\Gamma_{\ka}|^2}{N^2_{\text{at}}}
\sum_{\qu} 
 |\Gamma_{\qu}|^2 
 \sum_{\pe,\pe'}
n_{\pe} \tilde n_{\pe+ \ka}  \ n_{\pe'} \tilde n_{\pe'+ \qu} 
 \\
 \times
\Gf (\kappa_\ka,i\omega_{n} ) 
 \Gf (\kappa_\ka, i\omega_{n} ) , 
 \Gf (\kappa_\qu, i \omega_n + \omega_{\qu} - \omega_{\ka})
 \\
\times
\left(
1 +  \delta_{\pe, \pe'} + \delta_{\ka, \qu}+ \delta_{\pe + \ka, \pe' + \qu}
\right).
\end{gathered} \end{split}\end{equation}

\no
We introduce the effective interaction coupling parameter

\begin{equation}\begin{split} \begin{gathered}
F(\ka;\qu) =
\frac{1}{N^2_{\text{at}}} \sum_{\pe,\pe'}
\left(
1 +  \delta_{\pe, \pe'} + \delta_{\ka, \qu}+ \delta_{\pe + \ka, \pe' + \qu}
\right)
\\
\times
n_{\pe} \tilde n_{\pe+ \ka}  \ n_{\pe'} \tilde n_{\pe'+ \qu}, 
\end{gathered} \end{split} \end{equation}

\no
which simplifies the formula for the perturbed Green's function

\begin{equation}\begin{split}\begin{gathered}
\delta \Gf ^{(2)}(\kappa, i \omega_n)
=
 |\Gamma_{\ka}|^2 \, 
\Gf^3(\kappa, i \omega_n) 
\sum_{\qu}  |\Gamma_{\qu}|^2 \, 
F(\ka;\qu).
\end{gathered}\end{split}\end{equation}

\no
The function $F(\ka;\qu)$ can be calculated explicitly (see Appendix C).
Next step, we sum up over Matsubara frequencies to obtain the equal-time response

\begin{equation}\begin{split} \begin{gathered}
\Gf^{(2)}(\kappa,\tau = 0 )  
=
 |\Gamma_{\ka}|^2 
\sum_{\qu}  |\Gamma_{\qu}|^2
F(\ka;\qu) \
\frac{1} {\beta}
\sum_{i \omega_n}
\Gf^3(\kappa, i \omega_n) .
\end{gathered} \end{split}\end{equation}

\no
To proceed in the on-shell approximation, we use non-interacting propagators. 
The sum can be calculated using the standard Matsubara machinery \cite{Mahan}, yielding the result

\begin{equation}\begin{split} \begin{gathered}
\frac{1} {\beta}
\sum_{i \omega_n}
\Gf^3_0(\kappa, i \omega_n)  = 
\frac{\beta^2 e^{\beta \omega_{\kappa}}}{2} f^2_0(\omega_{\kappa})
- \beta^2 e^{2 \beta \omega_{\kappa}} f^3_0(\omega_{\kappa}),
\end{gathered} \end{split}\end{equation}

\no
where again $f_0(\omega) = \left\{ e^{\beta \omega} - 1 \right\}^{-1}$ is the Bose-Einstein distribution for photons. 
In the leading order, the perturbed Green's function is given by ($\mu=0$)

\begin{equation}\begin{split}\begin{gathered}
\label{perturbed nontrapped 2}
\Gf(\kappa,\tau = 0 )  = \Gf_0(\kappa,\tau = 0 ) + \beta e^{\beta \omega_{\kappa}}
\left(\Sigma^{(1)}_{\kappa} + \Sigma^{(2)}_{\kappa} \right) f^2_0(\omega_{\kappa})
\\
+{\mathcal{O}}\left[ f^3_0(\omega_{\kappa}) \right],
\end{gathered}\end{split}\end{equation}

\no
where the second self-energy contribution is determined as

\begin{equation}\begin{split}\begin{gathered}
\label{self-energy 2}
 \Sigma^{(2)}_{\kappa_\ka} 
 =
 \frac{\beta}{2} |\Gamma_\ka|^2 \sum_{\qu} |\Gamma_{\qu}|^2 F(\ka;\qu).
\end{gathered}\end{split}\end{equation}

\no
The formulas \eqref{perturbed nontrapped 2}-\eqref{self-energy 2} completely describe the renormalized equal-time response for the one-photon process ${\cal{H}}^{11}$ in the leading orders for a uniform system at $T \ne0$. 
The interactions thus modify the photon's spectrum,

\begin{equation}\begin{split}\begin{gathered}
\tilde \omega_{\kappa} = \omega_{\kappa}
+\Sigma^{(1)}_{\kappa} 
+\Sigma^{(2)}_{\kappa},
\ \ \ \ \
\omega_{\kappa} = \omega_0 + \frac{\kappa^2}{2 m^*}.
\end{gathered}\end{split}\end{equation}

\no
Including now the non-vanishing chemical potentials $\mu$, and keeping only linear terms in $\delta \mu =\mu - \mu_0 = \lim_{\kappa \to 0}[ \Sigma^{(1)}_{\kappa} 
+\Sigma^{(2)}_{\kappa} ]$, one obtains the distribution function of photons, modified by interactions, as

\begin{equation}\begin{split}\begin{gathered}
\label{47}
f(\tilde \omega_\kappa - \mu ) \approx 
f_0(\omega_\kappa  - \mu_0 ) 
\\
- 
 \beta  e^{\beta ( \omega_{\kappa} - \mu_0) }   f^2_0 (\omega_\kappa  - \mu_0)  
\left[
\Sigma^{(1)}_{\kappa} 
+\Sigma^{(2)}_{\kappa}  - \delta \mu \right] .
\end{gathered}\end{split}\end{equation}

\no
If integrated over $\kappa$, Eq.\eqref{47} gives the  critical number of photons at temperature $T$. 
Unfortunately, in one dimension the Bose-Einstein singularity is not integrable, so we need to modify the formalism by considering the  trapping potential.


\subsection{Perturbation theory in the trapping potential}

In this subsection, we calculate the renormalization of photon's propagators $\Gf$ as they are trapped in the microtube waveguide. This is done again by means of the perturbation theory. First, we look for the first corrections to the non-trapped propagator and then sum up corrections in all orders. This procedure gives a new propagator $G$, describing non-interacting but trapped photons. At the end of this subsection, this result is merged with the outcome of the previous subsection, giving the interacting trapped propagator in the leading order.

Consider non-interacting photons trapped in the potential $V(\ra)$ with its Fourier-image $V_{\ka}$. The Hamiltonian responsible for this reads

\begin{equation}\begin{split} \begin{gathered}
\label{48}
{\cal H} = 
\sum_{\ka} \omega_{\kappa_\ka} \, 
\phi^{\dag}_{\kappa_\ka} \phi^{}_{\kappa_\ka}
+
\sum_{\ka,\qu} V_{\qu}  \, 
\phi^{\dag}_{\kappa_{\ka+\qu}} \phi^{}_{\kappa_\ka}.
\end{gathered} \end{split}\end{equation}

\no
The perturbation theory is given by the same formalism [see \eqref{perturbation series}]. The renormalized photon propagator $G$ origins from the non-trapped propagator $G_0 \equiv \Gf$ and is augmented by the series of perturbative corrections,

\begin{equation}\begin{split} \begin{gathered}
\label{series trapped}
G(\kappa, \tau) = \Gf(\kappa, \tau) + \delta G^{(1)} (\kappa, \tau) + 
 \delta G^{(2)} (\kappa, \tau) + ... ,
\end{gathered} \end{split}\end{equation}

\begin{figure*}
	\includegraphics[width=1.0\textwidth]{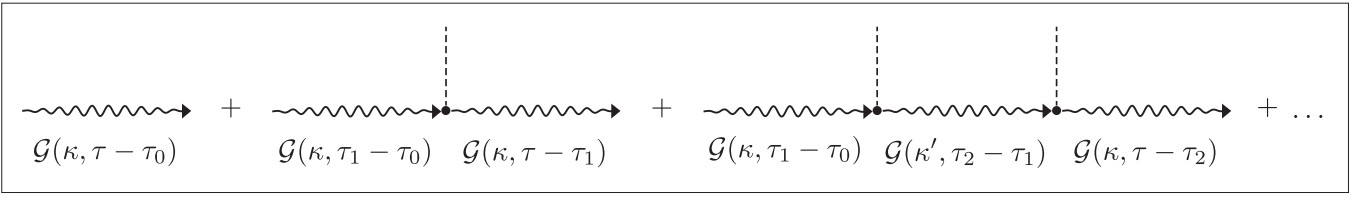}
	\caption{\label{trapped_propagator} Renormalization of a photon's propagator in the external field. Wavy propagators stand for one-dimensional photon Green's functions $\Gf (\kappa, \tau)$, whereas dashed lines denote interaction with external potential. 
The series converge up to the trapped photon propagator ${\text{G}}(\kappa,\tau )$. 	
	}
\end{figure*}

\no
as it is sketched in Fig.~\ref{trapped_propagator}. 
The first correction to the propagator due to interaction with the external potential (second diagram on Fig.~\ref{trapped_propagator}) is expressed as

%
%

\begin{equation}\begin{split}\begin{gathered}
\delta G^{(1)}(\kappa_\ka, \tau ) = 
 \sum_{\qu,\ka'} V_{\qu} 
 \int \limits_{0}^{\beta} d\tau_1 
 \\
 \times
 \langle  
\Torder   \, 
 \phi^{}_{\kappa_\ka}(\tau) \, 
 \phi^{\dag}_{\kappa_\ka}(0)  \, 
\phi^{\dag}_{\kappa_{\ka'+\qu}} (\tau_1) \, 
 \phi^{}_{\kappa_{\ka'}}  (\tau_1)
\rangle .
\end{gathered}\end{split}\end{equation}

\no
Using Wick's theorem, one obtains:

\begin{equation}\begin{split}\begin{gathered}
\delta G^{(1)}\left(\kappa_\ka, \tau \right)
= V_0  \int \limits_{0}^{\beta} d\tau_1  \, 
 \Gf \left(\kappa_\ka, \tau-\tau_1 \right) \, 
  \Gf \left( \kappa_\ka, \tau_1 \right) ,
\end{gathered}\end{split}\end{equation}

\no
where $V_0 \equiv V_{\kappa_\ka=0}$.
The convolution vanishes as we go to the Fourier transform, which gives

\begin{equation}\begin{split}\begin{gathered}
\delta G^{(1)}(\kappa, i \omega_n) = V_0  \ 
\Gf^2 (\kappa, i \omega_n) .
\end{gathered}\end{split}\end{equation}

\no
As in the previous calculations, the equal-time response is extracted from summing up over Matsubara frequencies, 

\begin{equation}\begin{split}\begin{gathered}
\delta G^{(1)}(\kappa, \tau = 0) = V_0  \ 
\frac{1}{\beta} \sum_{i \omega_n}
\Gf^2 (\kappa, i \omega_n)
= V_0 \, \frac {\beta e^{\beta \omega_{\kappa}}}{(e^{\beta \omega_{\kappa}}-1)^2} .
\end{gathered}\end{split}\end{equation}

\no
The long-wavelength asymptote $\kappa \to 0$ of the Fourier image $V_{\kappa}$ of potential $V(z)$ is finite and given by $V_0 = V(l)/(1+\alpha)$ (see Appendix B). 
The theoretical model presented in this section is not limited to $V(z) \propto |z|^{\alpha}$ with $\alpha = 1$, however for the sake of simplicity and beauty of the expressions we choose the linearly growing trapping potential, $\alpha = 1$, which gives

\begin{equation}\begin{split}\begin{gathered}
\label{54}
\delta G^{(1)}(\kappa, \tau = 0) = 
 \frac{e^{\beta \omega_k} f_0^2(\omega_{\kappa}) }{2 } \, \beta V(l) .
\end{gathered}\end{split}\end{equation}

The next correction is given by the third diagram in Fig.~\ref{trapped_propagator}, which is the only connected diagram one can build up with the perturbed Hamiltonian \eqref{48}. 
This diagram is expressed mathematically as

\begin{equation}\begin{split}\begin{gathered}
\delta G^{(2)}(\kappa_\ka, \tau ) =  
 \sum_{\qu} V_{\qu}  V_{-\qu}
\int \limits_{0}^{\beta} \int \limits_{0}^{\beta} 
d\tau_1   d\tau_2 \\
\times
\Gf (\kappa_\ka, \tau_1 )  \ 
\Gf(\kappa_{\ka+ \qu}, \tau_2 - \tau_1) \
 \Gf(\kappa_\ka, \tau - \tau_2), 
\end{gathered}\end{split}\end{equation}

\no
which upon Fourier transform yields to

\begin{equation}\begin{split}\begin{gathered}
\delta G^{(2)} (\kappa_\ka, i \omega_n  ) 
=  \Gf^2( \kappa_\ka, i \omega_n) 
 \sum_{\qu} | V_{\qu} |^2 
  \Gf( \kappa_{\ka+ \qu}, i \omega_n) . 
\end{gathered}\end{split}\end{equation}

\no
Therefore, the equal-time response is expressed in terms of the Matsubara frequency sum,

\begin{equation}\begin{split}\begin{gathered}
\delta G^{(2)} (\kappa_\ka,\tau = 0  ) =  \frac{1}{\beta} \sum_{\qu} | V_{\qu} |^2  \sum_{i \omega_n}
 \Gf^2(\kappa_\ka, i \omega_n)  
 \Gf(\kappa_{\ka+ \qu}, i \omega_n) 
 \\
 =
 \sum_{\qu} | V_{\qu} |^2  \  \frac{1}{\beta}  \sum_{i \omega_n}
 \frac{1}{(i \omega_n - \omega_{\kappa_\ka})^2 (i \omega_n - \omega_{\kappa_{\ka + \qu}})}.
\end{gathered}\end{split}\end{equation}

\no
Using the rules of summation, one obtains 
 the correction to the density of photons due to the interaction with external potential as

\begin{equation}\begin{split}\begin{gathered}
\label{58}
\delta G^{(2)}(\kappa_\ka,\tau = 0  ) 
=  
- \sum_{\qu} | V_{\qu} |^2  \\
\times  
 \left\{
  \frac{f_0(\omega_{\kappa_{\ka + \qu}}) - f_0(\omega_{\kappa_\ka}) }{(\omega_{\kappa_{\ka + \qu}} - \omega_{\kappa_\ka})^2}
 +
  \frac{\beta e^{\beta \omega_{\kappa_\ka}} \,  f_0^2(\omega_{\kappa_\ka}) }{ \omega_{\kappa_{\ka+\qu}} - \omega_{\kappa_\ka}}
 \right\}.
\end{gathered}\end{split}\end{equation}

\no
It's clear that the main contribution to the sum in Eq.\eqref{58} is given by the region where $\omega_{\kappa_{\ka + \qu}} \approx \omega_{\kappa_\ka}$. Therefore, we can expand the numerators in the Taylor series around $\omega_{\kappa_\ka}$. It is important to go up to the third order because some terms get canceled.
This in turn leads to

\begin{equation}\begin{split}\begin{gathered}
\delta G^{(2)}(\kappa,\tau = 0  ) =  
-  \frac{e^{\beta \omega_{\kappa}}(1+ e^{\beta \omega_{\kappa}} ) }{2}  \beta^2 f_0^3 (\omega_{\kappa}) 
\sum_{\qu} | V_{\qu} |^2  . 
\end{gathered}\end{split}\end{equation}

\no
Now one can calculate the sum here explicitly, which gives $\sum_{\qu} | V_{\qu} |^2   = V^2(l)/(1+2 \alpha)$ (see Appendix B). Taking again the linearly growing potential, $\alpha=1$,  one therefore obtains the expression

\begin{equation}\begin{split}\begin{gathered}
\label{60}
 \delta G^{(2)} (\kappa,\tau =0 )
  =  
-  \frac{e^{\beta \omega_{\kappa}}
(1+ e^{\beta \omega_{\kappa}} )
f_0^3 (\omega_{\kappa}) }{6}  \, \beta^2   V^2(l) ,
\end{gathered}\end{split}\end{equation}

\no 
which is the second-order correction to the free propagator due to the light trapping. Therefore, the first two corrections are given by Eq.\eqref{54} and Eq.\eqref{60}. 
For this study, it is important to go to the further orders. 
One can verify that in all orders the series \eqref{series trapped} converges to 

\begin{equation}\begin{split}\begin{gathered}
\label{trapped propagator}
G(\kappa, \tau = 0) = -
\frac{1}{\beta V(l)} \ln \frac{1- e^{\beta \omega_\kappa} e^{\beta V(l)} }
{(1- e^{\beta \omega_\kappa}) e^{\beta V(l)}},
\end{gathered}\end{split}\end{equation}

\no
which decribes the propagator in the external potential $V(z) = u |z|$. 
Again, the results can be obtained in the $l/R_0 \to \infty$ approximation and proceeding to the continuous spectrum by $\sum_{\ka} \to 2l \int \frac{d\kappa}{2 \pi}  $. 
Therefore, throughout the machinery of the previous derivations, one needs to replace the non-trapped propagators
$\Gf(\kappa, \tau)$ by the propagators $G(\kappa, \tau)$ of trapped photons. 
Up to the first order in photon-atom interactions  (i.e. neglecting all the higher-order terms), one obtains

\begin{equation}\begin{split}\begin{gathered}
\label{62}
\tilde f (\tilde \omega_{\kappa} - \mu) \approx
 -G(\kappa, \tau = 0) - \beta e^{\beta(\omega_{\kappa} - \mu)} 
 G^2(\kappa, \tau = 0) 
 \\
 \times
 \left(\Sigma_{\kappa} - \delta \mu \right) .
\end{gathered}\end{split}\end{equation}

\no
To calculate the critical photon number, one needs to sum up over momenta. 
However, a photon in the system under study is described by the motional degree of freedom $\kappa=k_z$, 
but there also could be
other degrees of freedom, for example, a polarizational degree of freedom. 
Recall that it was taken into account in formula \eqref{Ncrit} by introducing $\text{g}=\text{g}(k)$, which describes  degeneracy in photon energy levels. For the estimate in \eqref{Ncrit}, we took an effective $\text{g}^{*} \approx 3$, as a massive boson can exist in three polarization states even in on-shell approximation. Therefore, in this approximation the continuous limit is introduced as $\sum_{\ka} \to \frac{ \text{g} l}{\pi} \int_{-\infty}^{\infty} d \kappa$. This essentially leads to multiplying by $\text{g}^{*}$ each time we have wave-vector summation. Therefore, taking into account Eq.\eqref{62}, one obtains

\begin{equation}\begin{split}\begin{gathered}
\label{Ncrit_int}
N_{\phi} \approx \frac{2 \text{g}^{*} l }{\pi \beta V(l)} 
 \int \limits_{0}^{\infty} 
  d\kappa \,
   \ln
   \left[ \frac{1- e^{\beta (\omega_\kappa - \mu_0)} e^{\beta V(l)} }
{(1- e^{\beta (\omega_\kappa - \mu_0)}) \, e^{\beta V(l)}}
\right]
\\
-  \frac{2 \text{g}^{*} l }{\pi \beta V^2(l)}  
 \int \limits_{0}^{\infty}  d \kappa \,
 e^{\beta (\omega_{\kappa} - \mu_0)}
 \left( \Sigma_\kappa- \delta \mu \right) 
 \\
 \times
 \ln^{2} 
 \left[ \frac{1- e^{\beta (\omega_\kappa - \mu_0) } \,  e^{\beta V(l)} }
{(1- e^{\beta ( \omega_\kappa - \mu_0) }) \, e^{\beta V(l)}}
\right] .
\end{gathered}\end{split}\end{equation}

\no
One can verify -- either analytically or numerically --  that the first term is exactly the non-interacting result we discussed before [see the expression \eqref{N crit linear 1}].
Indeed, taking  $l \to \infty$, and $\mu_0 = \hbar \omega_0$, the first term turns into

\begin{equation}\begin{split}\begin{gathered}
N_{0} = \frac{2 \text{g}^{*} T}{\pi  u} 
 \int \limits_{0}^{\infty} 
  d\kappa \,
   \ln
   \left[1+ \frac{1} { \exp( \hbar^2 \kappa^2/2 m^{*} T)-1 }
\right]
\\
=
\int \limits_{-\infty}^{+\infty} \int \limits_{-\infty}^{+\infty} 
\frac{d \kappa d z}{2 \pi}
\frac{ \text{g}^{*}} { \exp\left[ 
\frac{\hbar^2 }{2 m^{*}T} \kappa^2 + \frac{u}{T}|z|
\right]- 1 }.
\end{gathered}\end{split}\end{equation}

\no
where $V(z) = u |z|$, i.e. $u = \hbar \omega_0/L_0$. 
One notices that it's exactly the formula \eqref{Ncrit} after relabelling motional degree of freedom $\kappa$ as $k_z$.


\subsection{Contributions from one-photon and two-photon processes}

Finally, in this subsection we list the contributions to self-energies from one-photon and two-photon processes \eqref{hierarchy}-\eqref{two-photon} without a detailed derivation. The derivation procedure is the same as for the process ${\cal{H}}^{11}$ in the two previous subsections. The critical number of photons in the interacting case involves 
the self-energies from the processes ${\cal{H}}^{11}$, ${\cal{H}}^{12}$, ${\cal{H}}^{13}$, ${\cal{H}}^{21}$, ${\cal{H}}^{22}$, ${\cal{H}}^{23}$, ${\cal{H}}^{24}$, ${\cal{H}}^{25}$, ${\cal{H}}^{26}$ which are given by

\begin{equation}\begin{split}\begin{gathered}
\Sigma_{\kappa} \approx 
\Sigma_{\phi} (\kappa) 
+\Sigma_{\phi \phi} (\kappa), 
\end{gathered}\end{split}\end{equation}

\no
where the number of ``$\phi$'' in the subscripts stands for the number of photons in an irreducible process, again in this study it is either one or two. The self-energies like $\Sigma_{\phi \phi \phi} (\kappa)$ and of higher orders are neglected.

The contributions from one-photon processes are given by

\begin{equation}\begin{split}\begin{gathered}
\label{self-energy one-photon}
\Sigma^{(1)}_{\phi}(\kappa_\ka) = 
 \sum_{\pe}
 \gamma^{11}_{\ka}  \, n_{\pe} \tilde n_{\pe + \ka}
  +
  \gamma^{12}_{\ka} \, n_{\pe}  n_{\pe + \ka}
  +
  \gamma^{13}_{\ka}  \, \tilde n_{\pe} \tilde n_{\pe + \ka}, 
\\
\Sigma^{(2)}_{\phi }(\kappa_{\ka}) = \frac{\beta}{2}  \sum_{\qu}
 \gamma^{11}_{\ka} \gamma^{11}_{\qu} F_1(\ka;\qu)
 +
  \gamma^{12}_{\ka} \gamma^{12}_{\qu} F_2(\ka;\qu)
  \\
 +
 \gamma^{13}_{\ka} \gamma^{13}_{\qu} F_3(\ka;\qu),
\end{gathered}\end{split}\end{equation}

\no
where $\gamma^{11}_{\ka} \approx |\Gamma^{11}_{\ka}|^2$, $\gamma^{12}_{\ka} = |\Gamma^{12}_{\ka}|^2$,
$\gamma^{13}_{\ka} = |\Gamma^{13}_{\ka}|^2$ are positive factors. In the main approximation the effective interaction parameters  (see Appendix C) are give by $F_{\alpha}(\ka;\qu) \approx \sigma_{\alpha} (\ka) \, \sigma_{\alpha} (\qu)$, where the quantities

\begin{equation}\begin{split}\begin{gathered}
\sigma_1(\ka) = 
\frac{1}{\Nat}
\sum_{\pe} n_{\pe} \tilde n_{\pe+\ka},
\\
\sigma_2(\ka) = 
\frac{1}{\Nat}
\sum_{\pe} n_{\pe}  n_{\pe+\ka},
\\ 
\sigma_3(\ka) = 
\frac{1}{\Nat}
\sum_{\pe} \tilde n_{\pe} \tilde n_{\pe+\ka},
\end{gathered}\end{split}\end{equation}

\no
can be calculated analytically (see Appendix C).

The contribution of the two-photon processes are given by

\begin{equation}\begin{split}\begin{gathered}
\label{self-energy two-photon}
\Sigma^{(1)}_{\phi \phi} (\kappa_{\ka})= 
\frac{1}{\Nat}
\sum_{\pe}
 r^{22}_{\ka} \,  n_{\pe} 
 + r^{23}_{\ka} \, \tilde n_{\pe}
 = r^{22}_{\ka} \frac{N_a}{\Nat}
  + r^{23}_{\ka} \frac{\tilde N_{a}}{\Nat},
\\
\Sigma^{(2)}_{\phi \phi}(\kappa) 
\approx  
 \frac{\beta}{2}
 \sum_{\qu} \gamma^{21}_{\ka\qu} \, \sigma_1 (\ka- \qu)
+\gamma^{22}_{\ka \qu}  \,  \sigma_2(\ka- \qu)
\\
+ \gamma^{23}_{\ka \qu} \,  \sigma_3 (\ka - \qu). 
\end{gathered}\end{split}\end{equation}

\no
where $r^{ab}_{\ka} =\Re \, \Gamma^{ab}_{\ka}$, and   $\gamma^{2 a}_{\ka \qu} = |\Gamma^{2 a}_{\ka}|^2$.
The expressions \eqref{self-energy one-photon}-\eqref{self-energy two-photon} describe the one-photon and two-photon processes without taking into account optical collisions of atoms (or molecules).

Therefore, the critical number of photons to observe the Bose-Einstein condensation at the temperature $T$ is defined by Eq.\eqref{Ncrit_int}, with the self-energies given by expressions \eqref{self-energy one-photon} - \eqref{self-energy two-photon}
and the renormalization of the non-interacting chemical potential, $\delta \mu = \lim _{\kappa \to 0} \Sigma_{\kappa}$.

An important simplification comes in the limit $l \to \infty$. In this case, the critical number of photons is given by

\begin{equation}\begin{split}\begin{gathered}
\label{69}
N_{C} \approx N_0
-  \frac{2 \text{g}^{*} L_0^2}{l} \frac{T}{\hbar \omega_0} 
 \int \limits_{0}^{\infty}  d \kappa 
\exp \left(\frac{\hbar^2 \kappa^2}{2 m^{*} T} \right)
 \frac{ \Sigma_\kappa - \delta \mu }{\hbar \omega_0}
 \\
 \times
\ln^{2} \left[
1+f_0 \left(\frac{\hbar^2 \kappa^2}{2 m^{*}T} \right)
\right] ,
\end{gathered}\end{split}\end{equation}

\no
where still $f_0 (x) = \left(e^x - 1\right)^{-1}$; the noninteracting critical number $N_0$ is given by formulas
\eqref{N crit linear 1} and \eqref{N crit linear 2}. 
Note that in the case $l \to \infty$ contributions from the lowest mean-field self-energies $\Sigma^{(1)}_{\phi}$ and $\Sigma^{(1)}_{\phi \phi}$ vanish, but contributions  from $\Sigma^{(2)}_{\phi}$ and $\Sigma^{(2)}_{\phi \phi}$ remain finite as they involve continuous-limit summation over photon momenta, returning the factor $l$.

%
%
%
%

\section{Discussion and Outline}

The main goal of this paper was to introduce the condensation of photons in one dimension: Find the necessary conditions, estimate the critical parameters, and look for the role of light-matter interactions.
However, it was not my goal to plan a particular experiment, and neither it was to plot the observables, since such calculations make sense only after (and only if) the experiment succeeds. 

	The analysis, presented in Sec.~\ref{Section1}, shows that in the weakly-interacting case, the condensation is possible if the light is trapped in a prolongated microtube, $l \gg R_0$, which is slowly narrowing towards the ends as a power-law function weaker than parabolic. The analysis has not been done for the strongly varying shape, $l \sim R_0$, as the quantization procedure in that case is not straightforward. However, I would not be surprised if a similar phenomenon, yet less distinct, could be observed for  $l \sim R_0$.
	
	The experiments on Bose-Einstein condensation of photons \cite{Klaers2010a,Klaers2010b,Klaers2011,Marelic2015} has an interesting distinguishing feature: the temperature of the setup is fixed, and the number of particles is tuned by external pumping. Therefore, one of the natural observables is $N_C(T)$ (compare to $T_C(N)$ in atomic BECs). In this study, the noninteracting model contributes the critical number of photons as $N_0 \propto T^{3/2}$, whereas the first perturbative corrections contribute in a more complicated manner (see formulas \eqref{Ncrit_int},\eqref{69} and the linked expressions in Appendix C).
Again, as in the  case of the two-dimensional BEC of photons \cite{Klaers2010b,Kruchkov2014}, 
the geometry of the system is important for tuning the system, since the parameters $R_0$, $l$, $L_0$ appear both in the non-interacting and interacting context.

The other interesting feature here is the sharp response to the atomic frequency resonance. As it was already mentioned, for the thermalization process based on the repeated processes of absorption and emission, it is important to ensure the closeness of the cut-off frequency $\omega_{0}$ and the main atomic transition frequency  $\omega_{\text{at}}$, so the absorption processes are favorable enough comparing to scattering processes. Even though the thermalization processes were not considered implicitly in this study, just referring to the earlier studies, the importance of the relation $\omega_{0} \approx \omega_{\text{at}}$ is apparent, as it appears throughout the paper, both in the noninteracting  and interacting cases:
The quantity $\Theta = e^{(\hbar \omega_0 - \hbar \omega_{\text{at}} )/T}$ reflects the  strength of this resonance for this system (see for example, Appendix C); also as the coupling parameters $\Gamma$ introduced in the Hamiltonian \eqref{hierarchy}-\eqref{two-photon} will have local extrema for momenta of photons, satisfying the relation $\hbar \omega( \ka) \simeq \hbar \omega_{\text{at}}$. 
Finally, for the completeness of this consideration, one should also add into account the average number of photons that are coupled with atoms, which also depends on the closeness to the atomic resonance. In equilibrium this quantity is linearly proportional to the number of atoms and is given by 
$N_{\text{at}}\left[ 1+ \text{g}_{12} \exp \left( \frac{\hbar \omega_{\text{at}}- \hbar \omega_0 }{T}\right)
\right]^{-1} $, where $\text{g}_{12}$ is the ratio between degeneracy of the atomic ground state and the first excited state, see for details Refs.\cite{Kruchkov2013,Kruchkov2014}.

The influence of indirect photon-photon interactions, mediated through the different processes of absorption, emission, and scattering, was studied in terms of an effective Hamiltonian, taking into account the hierarchy of multi-photon processes. Because the photon number in the system under study is significantly smaller than the number of atoms, the hierarchy graph can be truncated on the one-photon and two-photons processes, which give the leading contributions to self-energy, corresponding to the effectively Hartree-Fock terms if the direct photon-photon scattering was present. The temperature-dependent perturbation theory, represented here by Matsubara formalism, is valid in the symmetrical phase ($N \le N_C$), thus allowing us to calculate the critical parameters of the system. I should make here two important remarks. First, the different combinations of one-photon and two-photon processes can give interfering terms which, of course, will contribute to the self-energy, 
however this contribution appears to be significantly smaller;
 the lowest contributions of the three-photon processes 
  are of the higher order -- at least with eight photon operators, which is beyond the present study.
Second, there could also be be present different one-photon and two-photon processes, involving more than a pair of atoms, for example, optical collisions of a form 
$a^{\dag}   a^{\dag}  \phi^{\dag} a^{} \, a^{}$, $a^{\dag}   a^{\dag}  \phi^{\dag} \tilde a^{} \, \tilde a^{}$, etc. 
Even though formally these processes contribute into effectively Hartree-Fock decouplings, at least for the values of parameters used in the present paper the corresponding self-energies are negligible  comparing to the self-energy contributions given by formulas \eqref{self-energy one-photon} - \eqref{self-energy two-photon}. 

The problem, however, requires further study.  For example, 
for the photons in Bose-Einstein condensate, a more general formalism, allowing broken symmetry, is required.  A suitable machinery is given by the Popov approximation, --  I am currently working on it. It will be published elsewhere.

%
%
%
%

\appendix

\section{Atomic Green's functions}

The ground state atoms and excited state atoms are physically the same objects (although with different quantum numbers of electronic orbitals), so their propagators should be physically linked. On the other hand, in the model of two-level atoms, the dynamics of the system is described by two independent quantum operators $a_{\pe}$ (ground state) and $\tilde a_{\pe}$ (excited state).  
The generalized atomic operator is introduced as

\begin{equation}\begin{split}\begin{gathered}
A_{\pe} = 
\left( \begin{array}{c}
a_{\pe}  \\
\tilde a_{\pe}  \end{array} \right). 
\end{gathered}\end{split}\end{equation}

\no
One can verify that the number operator $A_{\pe}^{\dag} A_{\pe}$ returns the total number of atoms in the system,  - that is the number of atoms in the ground state $N_a$ together with the number of atoms in the excited state $N_{\tilde a}$, 

\begin{equation}\begin{split}\begin{gathered}
N_{\text{at}} = \sum_{\pe} A_{\pe}^{\dag} A_{\pe} 
= 
\sum_{\pe} a^{\dag}_{\pe}  a_{\pe} +  \tilde a^{\dag}_{\pe} \tilde a_{\pe}
=
N_a+N_{\tilde a} .
\end{gathered}\end{split}\end{equation}

\no
The generalized atomic Green's function is defined as follows

\begin{equation}\begin{split}\begin{gathered}
{\mathfrak{G} }  (\pe, \tau, \tau_0) =  - \left \langle 
\Torder \ 
A_{\pe}^{}  (\tau)
A_{\pe}^{\dag}  (\tau_0) 
\right \rangle_{\text{th}} .
\end{gathered}\end{split}\end{equation}

\no
Alternatively, it can also be presented in a matrix form,

\begin{equation}\begin{split}\begin{gathered}
{\mathfrak{G} } (\pe, \tau, \tau_0)=  
\left( \begin{array}{c c}
\Gf^{A}_{11} (\pe, \tau, \tau_0) \ 
 \Gf^{A}_{12} (\pe, \tau, \tau_0) \\
\Gf^{A}_{21} (\pe, \tau, \tau_0) \
 \Gf^{A}_{22} (\pe, \tau, \tau_0)
\end{array} \right)  ,
\end{gathered}\end{split}\end{equation}

\no
with the matrix elements given by

\begin{equation}\begin{split}\begin{gathered}
\Gf^{A}_{11} (\pe, \tau, \tau_0)  = - \left \langle  \Torder \ a_{\pe} (\tau) a^{\dag}_{\pe} (\tau_0) \right \rangle ,
\\
\Gf^{A}_{12} (\pe, \tau, \tau_0)  = - \left \langle  \Torder \ a_{\pe} (\tau)  \tilde a^{\dag}_{\pe} (\tau_0) \right \rangle ,
\\
\Gf^{A}_{21} (\pe, \tau, \tau_0)  = - \left \langle  \Torder \  \tilde a_{\pe} (\tau) a^{\dag}_{\pe} (\tau_0) \right \rangle ,
\\
\Gf^{A}_{22} (\pe, \tau, \tau_0)  = - \left \langle  \Torder \ \tilde a_{\pe} (\tau) \tilde a^{\dag}_{\pe} (\tau_0) \right \rangle.
\end{gathered}\end{split}\end{equation}

\no
The  off-diagonal propagators  compensate each other near equilibrium. Throughout the main text of the paper I use the notation 
$\Gfa(\pe,\tau) \equiv \Gf^{A}_{11} (\pe, \tau, 0)$ and
$ \Gfaa(\pe,\tau) \equiv \Gf^{A}_{22} (\pe, \tau, 0)$.

\section{Summation rules for potential harmonics}

The Fourier transforms $V_{\ka} = V_{\kappa_\ka}$ of the potential $V(z)$ are defined as

\begin{equation}\begin{split} \begin{gathered}
\label{Fourier transform}
V(z) = \sum_{\kappa} V_{\kappa} e^{i \kappa z},
\ \ \ 
V_{\kappa} = \frac{1}{2l} \int \limits_{-l}^{+l}dz \,  e^{- i \kappa z} V(z) .
\end{gathered} \end{split}\end{equation}

\no
One can verify by direct calculation  that the following property holds:

\begin{equation}\begin{split} \begin{gathered}
\label{sum}
\sum_{\kappa } V^2_{\kappa} = 
 \int  \limits_{-l}^{l} \frac{dz}{2l} \, V(z) V(-z) .
\end{gathered} \end{split}\end{equation}

\no
For symmetric potentials $V(z) = V(-z)$, thus one obtains

\begin{equation}\begin{split} \begin{gathered}
\sum_{\kappa } V^2_{\kappa} 
=   \frac{1}{l}  \int  \limits_{0}^{l} dz \, V^2(z) .
\end{gathered} \end{split}\end{equation}

\no
For the power-law models, $V(z) \propto |z|^\alpha$, 
one can carry out the integral explicitly as

\begin{equation}\begin{split} \begin{gathered}
\sum_{\kappa} V^2_{\kappa} 
=  \frac{V^2(z=l)}{1+2 \alpha}.
\end{gathered} \end{split}\end{equation}

\no
This sum remains finite for the important region $\alpha \in (0,2)$ and decreases monotonically,  approaching zero as $\alpha \to \infty$. 

We also need to calculate the  $\kappa=0$ term of this sum separately, which can be done as follows, 

\begin{equation}\begin{split} \begin{gathered}
V_{0}^2 = 
\left(
\frac{1}{2l} 
\int  \limits_{-l}^{l} dz    V(z)
\right)^2
=
\frac{V^2(z=l)}{(1+\alpha)^2},
\end{gathered} \end{split}\end{equation}

\no
where the last equality sign stands for $V(z) \propto |z|^\alpha$. 
Thus one the quantity $V_0 = V(l)/(1+\alpha)$ describes the long-scale physics of the problem.
Note that the main contribution to the sum \eqref{sum} is made by the low-energy photons, $ \omega \sim  \omega_0$.

\section{Effective interaction parameters}

The effective interaction parameters $F(\ka,\qu)$ determine the self-energy corrections to photon's spectrum, thus renormalizing the interacting vertices, reducing the photon-atoms-photon interaction to effective photon-photon interaction. Consider the first interaction parameter, given by

\begin{equation}\begin{split} \begin{gathered}
F_1(\ka;\qu) =
\frac{1}{N^2_{\text{at}}} \sum_{\pe,\pe'}
\left(
1 +  \delta_{\pe, \pe'} + \delta_{\ka, \qu}+ \delta_{\pe + \ka, \pe' + \qu}
\right)
\\
\times
n_{\pe} \tilde n_{\pe+ \ka}  \ n_{\pe'} \tilde n_{\pe'+ \qu}. 
\end{gathered} \end{split} \end{equation}

\no
Nevertheless all the sum can be calculated analytically, for the consistency of the main-order approximation the deltas should  be dropped, thus leading to

\begin{equation}\begin{split} \begin{gathered}
F_1(\ka;\qu) \approx
\frac{1}{N^2_{\text{at}}} \sum_{\pe,\pe'}
n_{\pe} \tilde n_{\pe+ \ka}  \ n_{\pe'} \tilde n_{\pe'+ \qu}, 
\end{gathered} \end{split} \end{equation}

\no
and the sum factorizes to

\begin{equation}\begin{split} \begin{gathered}
F_1(\ka;\qu) \approx
\sigma_1(\ka) \, \sigma_1(\qu), 
\ \ \ \ \ 
\sigma_1(\ka) = 
\frac{1}{\Nat}
\sum_{\pe} n_{\pe} \tilde n_{\pe + \ka} .
\end{gathered} \end{split} \end{equation}

\no
Therefore, we basically need to calculate $\sigma_{\ka}$. We make it in steps. First, let's calculate
the total number of atoms, which in nondegenerate system is given by

\begin{equation}\begin{split}\begin{gathered}
\Nat =
\sum_{\pe} n_{\pe} + \tilde n_{\pe}
=
  \sum_{\pe} 
e^{\beta(\mu_1 - E_{\pe})}+e^{\beta(\mu_2 - E_{\pe} - \Delta)},
\end{gathered}\end{split}\end{equation}

\no
where $\Delta = \hbar \omega_{\text{at}}$ is the energy difference between the two atomic eigenstates. The normalization is included as an additive constant to the chemical potentials. 
The chemical potentials $\mu_{1,2}$ of the ground state and excited state atoms are linked by  the condition of chemical equilibrium, 
 $\mu_1 + \mu = \mu_2$, with $\mu$ to be the chemical potential of photons. Because $\Nat$ is an observable, one can first derive the condition for the chemical potential for one of the atomic subsystem, 

\begin{equation}\begin{split}\begin{gathered}
\label{Theta}
e^{\beta \mu_1}  
=\frac{\Nat}{1+ \Theta}
\left( \sum_{\pe} e^{- \beta E_{\pe}} \right)^{-1},
\ \ \ \ \
\Theta(\mu) = e^{\beta (\mu - \Delta)},
\end{gathered}\end{split}\end{equation}

\no
and then calculate the effective interaction factor

\begin{equation}\begin{split}\begin{gathered}
\sigma_1(\ka) =
\frac{1}{\Nat}
 \sum_{\pe} n_{\pe} \tilde n_{\pe + \ka} 
\\
= 
\frac{1}{\Nat}
\sum_{\pe} 
e^{\beta(\mu_1 - E_{\pe})}
e^{\beta(\mu_2 - E_{\pe+\ka} - \Delta)} .
\end{gathered}\end{split}\end{equation}

\no
After this, upon the substitution of the previous formula \eqref{Theta}, and use of the energy-conservation relation 
 $E_{\pe} + \hbar \omega_{\ka} = E_{\pe+ \ka} + \Delta$, 
one obtains:

\begin{equation}\begin{split}\begin{gathered}
\sigma_1(\ka) = \frac{1}{2 \sqrt2}   \ e^{-\beta ( \hbar \omega_{\ka}-\mu)} 
\frac{\Nat}{(1+ \Theta)^2}
\left( \sum_{\pe} e^{- \beta E_{\pe}} \right)^{-1} .
\end{gathered}\end{split}\end{equation}

\no
The sum here can be calculated in the continuous approximation,

\begin{equation}\begin{split}\begin{gathered}
 \sum_{\pe} e^{- \beta E_{\pe}} 
 =
 \frac{ \cal{V} }{2 \sqrt{2} \pi^{3/2}} \left(\frac{m T}{\hbar^2} \right)^{3/2},
\end{gathered}\end{split}\end{equation}

\no
where ${\cal{V}}$ is the volume of the cavity with atoms. 
Therefore one obtains

\begin{equation}\begin{split}\begin{gathered}
\sigma_1(\ka) = \frac{\pi^{3/2} n_{\text{at}}}{(1 + \Theta)^2}
\left(\frac{\hbar^2}{m T} \right)^{3/2}
e^{- ( \hbar \omega_{\ka}-\mu)/T},
\end{gathered}\end{split}\end{equation}

\no
where $n_{\text{at}} =\Nat/ {\cal{V}}$ is the density of the atoms. 
One can verify that the quantity $\sigma_1(\ka)$ is dimensionless.

The similar expression can be obtained for the other effective interaction parameters $F(\ka,\qu)$. Again, in the main approximation they are given by

\begin{equation}\begin{split} \begin{gathered}
F_2(\ka;\qu) \approx
\sigma_2(\ka) \, \sigma_2(\qu), 
\ \ \ \ \ 
\sigma_2(\ka) = 
\frac{1}{\Nat}
\sum_{\pe} n_{\pe}  n_{\pe + \ka} ,
\\
F_3(\ka;\qu) \approx
\sigma_3(\ka) \, \sigma_3(\qu), 
\ \ \ \ \ 
\sigma_3(\ka) = 
\frac{1}{\Nat}
\sum_{\pe} \tilde n_{\pe} \tilde n_{\pe + \ka}.
\end{gathered} \end{split} \end{equation}

\no
The effective interaction factors are calculated within the same approximations and are given by

\begin{equation}\begin{split}\begin{gathered}
\sigma_2(\ka) = \frac{\pi^{3/2} }{(1 + \Theta)^2}
n_{\text{at}}
\left(\frac{\hbar^2}{m T} \right)^{3/2}
e^{- ( \hbar \omega_{\ka}-\Delta)/T},
\\
\sigma_3(\ka) = \frac{\pi^{3/2} \Theta}{(1 + \Theta)^2} n_{\text{at}}
\left(\frac{\hbar^2}{m T} \right)^{3/2}
e^{- ( \hbar \omega_{\ka}-\mu)/T}.
\end{gathered}\end{split}\end{equation}

\no
Finally, one can notice that
$\sigma_2(\ka) = \sigma_1(\ka) / \Theta$ 
and 
$\sigma_3(\ka) = \sigma_1(\ka) \,  \Theta$,
where $\Theta$ is given in formula \eqref{Theta}. 
As  $\Theta$  depends on $\mu$ that in the leading order is  $\hbar \omega_0$ in the case of condensate, therefore changing the geometry of the cavity one can make certain  processes to be more or less important for the system under consideration.

%
%
%
%

\end{document}